\documentclass[a4paper,10pt]{article}
\usepackage[utf8]{inputenc}
\usepackage{fancybox} 
\usepackage{tabularx} 
\usepackage{multirow}
\usepackage{multicol}
\usepackage{booktabs}
\usepackage{grffile}
\usepackage[dvips]{graphicx}

\newcommand{\lsim}{\raisebox{-4pt}{$\,\stackrel{\textstyle
                                                         <}{\sim}\,$}}

\newcommand{\nn}{\nonumber}
\newcommand{\be}{\begin{equation}}
\newcommand{\ee}{\end{equation}}
\newcommand{\ba}{\begin{eqnarray}}
\newcommand{\ea}{\end{eqnarray}}
\newcommand{\req}[1]{(\ref{#1})}
\def\={\,=\,}
\newcommand{\ci}[1]{\cite{#1}}

\def\mev{~{\rm MeV}}
\def\gev{~{\rm GeV}}

\def\ale{\alpha_{\rm em}}
\def\als{\alpha_{\rm s}}
\def\eps{\epsilon}

\newcommand{\tw}{\textwidth}

\def\vb0{{\bf b}_0}

\newcommand{\ov}[1]{\overline#1}

\def\={\,=\,}

\begin{document} 
\thispagestyle{empty}
\thispagestyle{empty}
\begin{flushright}
August, 31  2020\\[20mm]
\end{flushright}

\begin{center}

\vskip 10mm
{\Large\bf Lepton-pair production in hard exclusive hadron-hadron collisions} \\
\vskip 4mm

S.V.\ Goloskokov$\,^\S$~\hspace*{-0.01\tw}\footnote{Email:  goloskkv@theor.jinr.ru},
P.\ Kroll$\,^\dag$~\hspace*{-0.01\tw}\footnote{Email:  kroll@physik.uni-wuppertal.de} and
O.\ Teryaev$\,^{\S\ddagger}$~\hspace*{-0.01\tw}\footnote{Email: teryaev@jinr.ru}\\ [2em]

\S:\, {\small {\it Bogoliubov Laboratory of Theoretical Physics, Joint Institute
for Nuclear Research, Dubna 141980, Moscow region, Russia}}\\
\dag:\, {\small {\it Fachbereich Physik, Universit\"at Wuppertal, D-42097 Wuppertal,
    Germany}}\\
$\ddagger:$\, {\small {\it Veksler and Baldin  Laboratory of High Energy Physics,
            Dubna 141980,\\ Moscow region, Russia}}
\end{center}

\vskip 5mm 
\begin{abstract}
  \noindent We investigate lepton-pair production in hard exclusive hadron-hadron collisions.
  We consider a double handbag (DH) mechanism in which the process amplitude factorizes in hard
  subprocesses, $qq\to qq\gamma^*$ and $qg\to qg\gamma^*$, and in soft hadron matrix elements
  parameterized as generalized parton distributions (GPDs). Employing GPDs extracted from exclusive
  meson electroproduction, we present predictions for the lepton-pair cross section at kinematics
  typical for the LHC, NICA and FAIR. It turns out from our numerical studies that the
  quark-gluon subprocess dominates by far, the quark-quark(antiquark) subprocesses are almost negligible.
\end{abstract}   

\section{Introduction}
In the last two decades there were a lot of activities in the measurement and the theoretical
analysis of hard exclusive processes, such as deeply virtual Compton scattering or
electroproduction of mesons. The theoretical analyses of these processes, beginning with the
pioneering articles by Ji \ci{ji96} and Radyushkin \ci{rad96}, bases on factorization of the
process amplitudes in hard, perturbatively calculable parton-level subprocesses and in soft
hadronic matrix elements, parameterized as general parton distributions (GPDs). This type
of factorization, often dubbed as the handbag approach, has been shown to hold
for the mentioned processes \ci{collins96,collins98} in the generalized Bjorken regime of large photon
virtuality, $Q^2$, and large center-of-mass energy at fixed Bjorken-$x$ and small squared
invariant momentum transfer, $t$ ($-t\ll Q^2$). This type of factorization has also been applied
to wide-angle processes, e.g.\ \ci{rad98,DFJK1}, and time-like ones, e.g.\ \ci{diehl98}. In contrast
to the deeply virtual processes,  rigorous proofs of factorization do not exist for the latter
processes. Experimental and theoretical investigations,
however, revealed that for moderately large photon virtualities there are frequently substantial
corrections to the asymptotic handbag picture, e.g.\ in $\pi^0$ electroproduction \ci{dufurne,GK6}.

\begin{figure}[t]
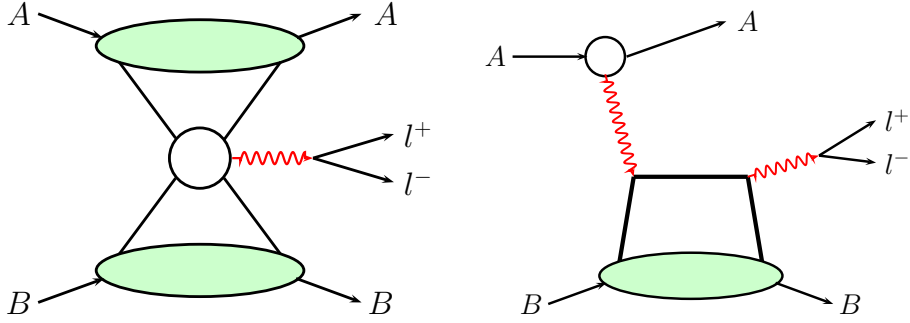

\centering
\includegraphics[width=0.46\tw]{double-handbag.epsi} \hspace*{0.04\tw}
\includegraphics[width=0.46\tw]{TVCS.epsi}
\caption{\label{fig:double-handbag} Left: The double handbag for exclusive lepton-pair
production in hadron-dadron collisions. Right: The TVCS mechanism.}
\end{figure}

In the present paper we are interested in lepton-pair production in hard exclusive hadronic collisions
\be
A\; B \to A\; B\; l^+\; l^-
\label{eq:process}
\ee
at large Mandelstam $s$ and large, time-like photon virtualities but small momentum transfer.
This exclusive analogue of the Drell-Yan process can, in principle, be measured at the LHC and at the future
accelerators NICA, FAIR and J-PARC.
We assume the above described factorization in hard parton-level subprocesses and in soft proton matrix
elements to hold and describe the process \req{eq:process} by a double handbag, see Fig.\ \ref{fig:double-handbag}.
The double handbag is also appearing due to analytic properties of the relevant amplitude \cite{teryaev:2005uj}).
The DH mechanism has already been applied to exclusive reactions involving charmed hadrons such as
$p\bar{p}\to \Lambda_c\ov{\Lambda}_c$ \ci{schweiger09} or $\pi^-p\to D^-\Lambda_c^+$ \ci{kofler15}. In
these reactions the large scale is set by the mass of the charm quark. An alternative dynamical
mechanism to the double handbag is depicted on the right-hand side of Fig.\ \ref{fig:double-handbag}:
A photon is emitted from one of the hadrons and interacts with a constituent quark from the other hadron
in the sense of time-like virtual Compton scattering (TVCS). This process has been studied theoretically, e.g.\
in \ci{DPB,moutarde19} but has not yet been measured. We expect that this single-handbag mechanism leads to much
smaller cross sections than the double handbag except perhaps in ultraperipheral heavy ion reactions. The strong quark-gluon
subprocesses contributing only to the double handbag will dominate as we are going to demonstrate in the following
sections. The single-handbag mechanism has been advocated for by Cisek et al \ci{cisek} for the production
of heavy vector mesons, like the $J/\Psi$, in semiexclusive hadron-hadron and hadron-nucleus collisions.
Of course, the purely electromagnetic lepton-pair production is also to be considered by us.\\

The plan of the paper is as follows:
A  kinematical prelude is  presented in the next section and, in Sect.\ \ref{sec:subprocess},  the hard
subprocesses are described in some detail. The full $A\; B \to A\; B\; l^+\; l^-$ amplitudes, given as
convolutions of the subprocess amplitudes and GPDs, are discussed in Sect.\ \ref{sec:amplitudes}.
In the following section the purely electromagnetic generation of lepton pairs in exclusive hadronic
collisions is discussed. The DH amplitudes, derived in Sect.\ \ref{sec:amplitudes},
are specified for particular processes in Sect.\ \ref{sec:results} and some numerical results
for cross sections are given. We also discuss the relative strength of the electromagnetic and DH
contributions in this section. In the Appendix some useful formulas for the phase space and the decay of the virtual
photon are repeated.  

\section{Kinematics}
\label{sec:kinematics}
First we consider the process
$A(p_a,\mu_a)\;B(p_b,\mu_b) \to A(q_1,\mu_1)\;B(q_2,\mu_2)\;\gamma^*(q_3,\nu)$ where $A$ and $B$
are protons, antiprotons or pions.  The decay of the virtual photon into the lepton pair will be
treated separately in App.\ A. The momenta and helicities of the various particles are denoted
by $p_i,q_i$ and $\mu_a,\ldots \mu_2$ and $\nu$, respectively.
Since we assume large $s$ the hadron masses will be neglected throughout the paper.
Momentum conservation tells us that
\be
p_a + p_b \= q_1 + q_2 + q_3\,.
\ee
It is convenient to introduce the following Lorentz invariants in addition to $s=(p_a+p_b)^2$
\ba
s_1&=& (q_1+q_3)^2\,,  \qquad s_2\=(q_2+q_3)^2\,, \nn\\
t_1&=& (p_a-q_1)^2\,,  \qquad t_2\=(p_b-q_2)^2\,.
\ea
The following relations hold \ci{kajantie}
\ba
s_1+t_1-t_2 &=& 2p_a\cdot q_3\,, \nn\\
s_2+t_2-t_1 &=& 2 p_b\cdot q_3\,,
\ea
and, in the limit $t_1, t_2\to0$,
\be
s_1\simeq s_2 \simeq \sqrt{s} Q\,.
\label{eq:forward}
\ee
The hadron momenta, given in light-cone coordinates, are parameterized in Ji's frame \ci{ji98} by:
\ba
p_a &=& \Big[(1+\xi_1)\bar{p}^+_a, \frac{\Delta^2_{1\perp}}{8(1+\xi_1)\bar{p}^+_a},
                                                                -\frac12\Delta_{1\perp},0\Big]\,, \nn\\
p_b &=& \Big[ \frac{\Delta^2_{1\perp}}{8(1+\xi_2)\bar{p}^-_b},   (1+\xi_2)\bar{p}^-_b,
                                                          \phantom{-} \frac12\Delta_{1\perp},0\Big]\,, \nn\\
q_1&=& \Big[(1-\xi_1)\bar{p}^+_a, \frac{\Delta^2_{1\perp}}{8(1-\xi_1)\bar{p}^+_a},
                                                           \phantom{-} \frac12\Delta_{1\perp},0\Big]\,, \nn\\
q_2&=& \Big[ \frac{\Delta^2_{2\perp}}{8(1-\xi_2)\bar{p}^-_b},   (1-\xi_2)\bar{p}^-_b,
                                                                   -\frac12\Delta_{2\perp}\cos{\Phi_2},
                                                               -\frac12\Delta_{2\perp}\sin{\Phi_2}\Big]\,,
\ea
where $\bar{p}_a^+$ ($\bar{p}_b^-$) is the plus (minus) component of the average hadron momentum
$\bar{p}_a=(p_a+q_1)/2$ ($\bar{p}_b=(p_b+q_2)/2$) at the upper (lower) hadronic blob of the graph shown
on the left-hand side of Fig.\ \ref{fig:double-handbag}. As usual the skewness parameter, $\xi_1$ ($\xi_2$),
represents the ratio of the plus (minus) components of the difference and the sum of the hadron momenta at
the upper (lower) blob. The skewness parameters as well as $\bar{p}_a^+$ and $\bar{p}_b^-$ can be expressed
in terms of the invariants. Up to corrections of order $t_i/Q^2$ ($i=1,2$), the expressions read
\ba
\bar{p}_a^+&=& \frac{2s-s_2}{2\sqrt{2s}}\,,  \qquad
                       \bar{p}_b^-\=\frac{\sqrt{s}}{2\sqrt{2}}\,\frac{4s-2s_1-2s_2+Q^2}{2s-s_2}\,,\nn\\
\xi_1&=&\frac{s_2}{2s-s_2}\,,  \qquad \xi_2\=\frac{2s_1-Q^2}{4s-2s_1-2s_2+Q^2}\,.
\label{eq:skewness}
\ea
Working in that frame means that we have to use the GPDs according to Ji's definition \ci{ji98}. These GPDs are
invariant under boosts in the 3-direction and under rotations around the 3-axis but are not invariant, for
instance, under rotations around the 2-axis.

In Ji's frame the Mandelstam $t_i$ are related to the momentum transfers, $\Delta_{i\perp}$, by
\ba
t_1&=&-\frac{\Delta_{1\perp}^2}{1-\xi_1^2}\,,    \nn\\
t_2&=&-\frac14\,\frac{\Big[(1-\xi_2)\,\Delta_{1\perp} - (1+\xi_2)\,\Delta_{2\perp}\Big]^2}{1-\xi_2^2}
                           - \Delta_{1\perp}\Delta_{2\perp}\,\sin^2{(\Phi_2/2)}\,.
\ea
Thus, the limits $t_i\to 0$ imply $\Delta_{i\perp}\to 0$.
\section{The subprocess amplitudes}
\label{sec:subprocess}
As already mentioned we are interested in the process $AB\to AB\gamma^*$ at large $s$, large photon virtuality,
$q_3^2=Q^2$, but small momentum transfer at the hadronic vertices, $t_i\ll Q^2$ ($i=1,2$). 
According to the handbag factorization it is assumed that partons are emitted and reabsorbed from the hadronic
blobs collinear to the hadronic momenta. Since in the hard subprocesses there are no soft parameters available
all dimension full variables have to be scaled by the hard scale, the photon virtuality, $Q^2$. Due to our
supposition of $t_i\ll Q^2$ we have to calculate the subprocess amplitudes in the limit $t_i\to 0$. In this
limit the parton momenta simplify to
\ba
k_a&=&\big[(x_1+\xi_1)\bar{p}_a^+\,,0\,,{\bf 0_\perp}\,\big]\,,  \qquad
                                                k_b\=\big[0\,,(x_2+\xi_2)\bar{p}_b^-\,,{\bf 0_\perp}\,\big]\,, \nn\\
k_1&=&\big[(x_1-\xi_1)\bar{p}_a^+\,,0\,,{\bf 0_\perp}\,\big]\,,  \qquad
                                                k_2\=\big[0\,,(x_2-\xi_2)\bar{p}_b^-\,,{\bf 0_\perp}\,\big]\,.
\ea
We see that the parton momenta attached to the upper vertex in Fig.\ \ref{fig:double-handbag} have large plus
components whereas those emitted and reabsorbed from the lower vertex have large minus components.
In terms of the parton momenta the virtual photon momentum is approximately given by
\be
q_3\simeq \big[2\xi_1\bar{p}_a^+\,,2\xi_2\bar{p}_b^-\,,{\bf 0_\perp}\,\big]\,.
\ee
With the help of \req{eq:forward} and \req{eq:skewness} one readily sees that we correctly have 
\be
q_3^2\simeq Q^2\,.
\ee

\begin{figure}[t]
\centering
  \includegraphics[width=0.60\tw]{LO-qg-graphs.epsi}
\caption{\label{fig:graphs-qg} Typical leading order Feynman graphs for
  the subprocesses $qg\to qg\gamma^*$.}
\end{figure}
\begin{figure}[t]
\centering
  \includegraphics[width=0.60\tw]{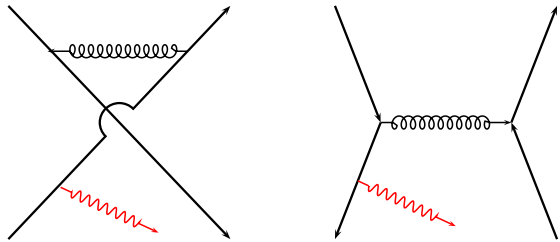}
\caption{\label{fig:graphs-qq} Typical leading order Feynman graphs for
  the subprocess $qq\to qq\gamma^*$.}
\end{figure}

 
We are going to compute the subprocess amplitudes to leading-order of QCD and to leading-twist accuracy. 
The possible subrocesses are $q(\bar{q})g\to q(\bar{q})g\gamma^*$ and $qq(\bar{q})\to qq(\bar{q})\gamma^*$.
The subprocess $gg\to gg\gamma^*$ is suppressed by $\als$ and, hence, neglected. Typical leading order
Feynman graphs for the relevant subprocesses are shown in Figs.\ \ref{fig:graphs-qg} and \ref{fig:graphs-qq}.
Obviously, the virtual photon is to be coupled to all quark lines. Since the parton helicities are not observed they
have to be averaged over with some projector operator onto the hadronic state. Thus, in the following, we will deal
with the amplitude for the subprocess $bc\to bc \gamma^*$ summed over the helicities, $\lambda_b$, $\lambda_c$, of the
partons $b$ and $c$ which corresponds to the projector $p_\mu  \gamma^\mu$ (being also the density matrix of unpolarized
partons, when the imaginary part of the amplitude is considered by the use of the optical theorem): 
 \be
 {\cal H}^{bc}_\nu\= \frac14\,\sum_{\lambda_b\lambda_c} {\cal H}^{bc}_{\lambda_b\lambda_c\nu,\lambda_b\lambda_c}\,.
\label{eq:sub-amp}
 \ee
 Since we are dealing with light quarks any quark-helicity-flip amplitude is zero.
 Nevertheless, the quarks or antiquarks emitted and reabsorbed from a hadronic vertex may have opposite
 helicities. Such configurations come from subprocesses like $q(+)q(-)\to q(-)q(+)\gamma^*$ or
 $q(+)\bar{q}(-)\to q(-)\bar{q}(+)\gamma^*$. In these cases the corresponding subprocess amplitudes are to be
 convoluted with transversity GPDs. Contributions of this type are neglected in this work. It is expected that for
 valence quarks this contribution is of about the same magnitude as the contribution from the valence-quark
 GPD $H$ \ci{GK6,GK7}. The gluonic transversity GPDs do not contribute here since the corresponding subprocess
 amplitude vanishes for $t_i\to 0$. As our numerical analysis reveals, see Sect.\ \ref{sec:results}, the
 dominant contribution to the processes of interest comes from the quark-gluon subprocess.
 
 Straightforward calculations of the subprocesses reveal that only those for longitudinal polarization of the
 virtual photon ($\nu=0$) are non-zero which is natural because of the similarity of the amplitude under consideration to
 the mesonic formfactors \cite{pivovarov}, where distribution amplitudes enter instead of  GPDs.
 For the quark-gluon subprocess, see Fig.\ \ref{fig:graphs-qg} for the relevant Feynman graphs, the non-zero
 amplitude reads~\footnote{
        In addition to the momentum-space Feynman expressions there is a factor $1/\sqrt{x_1^2-\xi_1^2}$ coming
        from the initial integration over $k^-$:
        $$\frac{d\bar{k}^-}{\sqrt{4k_a^-k_1^-}}=\frac{dx_1}{\sqrt{x_1^2-\xi_1^2}} $$
        and, as a consequence of the use of light-cone gauge, a factor $1/\big[(x_2+\xi_2)(x_2-\xi_2+i\eps)\big]$
        arising from converting the gluon field, $A_\mu$, appearing in the perturbative calculation, into the gluon
        field strength tensor, $G_{\mu\nu}$, in terms of which the gluonic GPDs are defined \ci{rad96,kogut-soper}.
        The latter factor is absorbed in the subprocess amplitudes.}
\ba
   {\cal H}_{0}^{qg}(x_1,\xi_1,x_2,\xi_2)&=& -16\pi  \frac{\als(Q^2)}{N_cQ} \,\frac1{x_2+\xi_2}\nn\\
           &\times& \left\{
                   \Big[\frac1{x_1-\xi_1-i\eps} - \frac1{x_1+\xi_1+i\eps}\Big]\,\frac1{x_2-\xi_2+i\eps} \right. \nn\\
                   &+& \left. \frac{\xi_1}{x_1+\xi_1+i\eps} \frac{1}{(x_1-\xi_1)(x_2-\xi_2)+i\eps} \right.\nn\\
                   &+& \left. \frac{\xi_1}{x_1-\xi_1-i\eps} \frac{1}{(x_1+\xi_1)(x_2-\xi_2)-i\eps} \right\}\,.
\label{eq:amplitude-qg}
\ea
The momentum fraction $x_2$ as well as  the skewness $\xi_2$ refer to the gluon. For the amplitude ${\cal H}^{gq}_{0}$
one has to interchange $x_1$ and $x_2$ as well as $\xi_1$ and $\xi_2$. In \req{eq:amplitude-qg} $N_c$ is the number
of colors. The QCD coupling constant, $\als$, is evaluated at the hard scale, the
photon virtuality $Q^2$, from the one-loop expression with $\Lambda_{\rm QCD}=220\,\mev$ and three flavors.
Since the integral over the gluon's
momentum fraction, $x_2$, extends only from 0 to 1 as a consequence of the fact that the gluon GPD $H$ is
an even function of $x$, we have to consider only positive values of $x_2$. Hence, the term $1/(x_2+\xi_2)$ is
not singular. The most singular integrals come from the terms
\be
        \frac1{(x_1 \pm \xi_1)(x_2\ - \xi_2) \mp i\eps}
\ee
which is approximated by
\be
\frac1{(x_1 \pm \xi_1)(x_2\ - \xi_2) \mp i\eps} \simeq \frac1{x_1 \pm \xi_1 \mp i\eps}\, \frac1{x_2 - \xi_2 \mp i\eps}\,.
\label{eq:recipe}
\ee
  The recipe \req{eq:recipe} can be justified to some extent by starting from the unphysical region with
  $|\xi_i| > 1$ and perform an analytic continuation \cite{teryaev:2005uj} to the physical region where
  $|\xi_i| < 1$. The remaining problem is the  different sign of $\eps$ for the continuation in $s$ and
  $s_1, s_2$. There are arguments \cite{Teryaev:2010mpa} that the double spectral representation and a
  symmetric continuation in $s_1, s_2$ should be considered corresponding to positive $\eps$. A special
  role is played by the inapplicability of the Steinmann relation because a virtual photon is involved
  in the process of interest. The interference with the electromagnetic contribution is especially
  interesting as only the real part of the DH amplitude is entering which is insensitive to the
  mentioned sign.  

From \req{eq:amplitude-qg} one sees that ${\cal H}_0^{qg}$ possess the property 
\be
   {\cal H}_{0}^{qg}(-x_1,\xi_1,x_2,\xi_2)\={\cal H}_{0}^{qg}(x_1,\xi_1,x_2,\xi_2)\,.
\label{eq:property-qg}
\ee

The $qq\to qq \gamma^*$ subprocess amplitude is to be computed analogously from the leading-order
Feynman graphs of the type shown on the left-hand side of Fig. \ref{fig:graphs-qq}:
\ba
   {\cal H}_{0}^{qq}(x_1,\xi_1,x_2,\xi_2) &=& - 16\pi \frac{\als(Q^2)}{N_cQ} \,
   \left\{\frac1{(x_1-\xi_1)(x_2+\xi_2)-i\eps} \right. \nn\\
      && \left.\hspace*{0.1\tw}  - \frac1{(x_1+\xi_1)(x_2-\xi_2)-i\eps} \right\} \,.
\label{eq:amplitude-qq}
\ea
The singular integrals are regularized according to Eq.\ \req{eq:recipe}.
From \req{eq:amplitude-qq} the following symmetry properties are evident 
\ba
{\cal H}_{0}^{qq}(-x_1,\xi_1,-x_2,\xi_2) &=& - {\cal H}_{0}^{qq}(x_1,\xi_1,x_2,\xi_2)\,,  \nn\\
{\cal H}_{0}^{qq}(-x_1,\;\xi_1,\;x_2,\;\xi_2) &=& - {\cal H}_{0}^{qq}(x_1,\xi_1,-x_2,\xi_2) \nn\\
                          &=& - 16\pi \frac{\als(Q^2)}{N_cQ} \,
                               \left\{\frac1{(x_1-\xi_1)(x_2-\xi_2)+i\eps}  \right. \nn\\ 
                       &&\left.\hspace*{0.18\tw} - \frac1{(x_1+\xi_1)(x_2+\xi_2)+i\eps} \right\}\,.
\label{eq:rel-H}
\ea
With the help of these symmetry relations one finds for the $\bar{q}\bar{q}\to \bar{q}\bar{q}\gamma^*$
amplitude
\be
   {\cal H}_{0}^{\bar{q}\bar{q}}(x_1,\xi_1,x_2,\xi_2)\= {\cal H}_{0}^{qq}(-x_1,\xi_1,-x_2,\xi_2)
                               \=- {\cal H}_{0}^{qq}(x_1,\xi_1,x_2,\xi_2)
\ee
and for the $q\bar{q}$ ($\bar{q}q$) one which is to be calculated from the graphs shown on the
right-hand side of Fig.\ \ref{fig:graphs-qq},
\be
   {\cal H}_{0}^{q\bar{q}}(x_1,\xi_1,x_2,\xi_2) \=  {\cal H}_{0}^{\bar{q}q}(x_1,\xi_1,x_2,\xi_2)
                                             \=  {\cal H}_{0}^{qq}(x_1,\xi_1,-x_2,\xi_2)\,.
\ee
As it becomes clear in the following section both the types of graphs shown in Fig.\ \ref{fig:graphs-qq}
lead to the same convolutions. Therefore, only the subprocess amplitude ${\cal H}_{0}^{qq}$, defined in
Eq.\ \req{eq:amplitude-qq}, is to be taken into account
because the $\bar{q}\bar{q}$, $q\bar{q}$ and $\bar{q}q$ amplitudes are contained in the convolution of
${\cal H}^{qq}_0$ with a relevant GPD implying integrations over $x_i$ from -1 to 1.\\

\section{The process amplitudes}
\label{sec:amplitudes}
In the kinematical domain of interest in the present work helicity non-flip vertices $A\to A$ and $B\to B$
(i.e\ $\mu_a=\mu_1$, $\mu_b=\mu_2$) dominate which, in the handbag approach, are under control of
the GPD $H$. For proton and antiproton $H$ is strictly speaking the combination
\be
H_{\rm eff}\= H - \frac{\xi_i^2}{1-\xi_i^2} E\,. 
\ee
Contributions from the other GPDs like $\widetilde{H}$, $E$ or from transversity
GPDs are expected to be small according to experience with GPDs extracted from data on electroproduction
of vector mesons \ci{GK3,status}. Hence, we approximate $H_{\rm eff}$ by $H$.  The full process amplitudes
are given by the  convolution~\footnote{
         For a pion vertex the factor $\sqrt{1-\xi_i^2}$ does not appear.}
\ba
   {\cal M}^{AB}_{\mu_a\mu_b\nu,\mu_a\mu_b} &=& e_0 \sqrt{1-\xi_1^2} \sqrt{1-\xi_2^2}\, \sum_{a=u,d,s} e_a
                          \sum_{b,c} \int dx_1 dx_2 \nn\\
                    &\times&  H_A^b(x_1,\xi_1,t_1) H_B^c(x_2,\xi_2,t_2)\,{\cal H}_{0}^{bc}(x_1,\xi_1,x_2,\xi_2)
\label{eq:amplitude-AB}
 \ea
 where $b$ and $c$ are either a quark of flavor $a$ or a gluon, $H_A^b(x_1,\xi_1,t_1)$ ($H_B^c(x_2,\xi_2,t_2)$)
 is a quark or gluon GPD of the hadron $A$ ($B$). The variable $x_1(x_2)$ is the average momentum fraction
 at the vertex $A\to A$ ($B\to B$). The positron charge is denoted by $e_0$ and $e_a$ the charge of the
 quark in units of $e_0$. Last not least, ${\cal H}^{bc}_{0}$ is the subprocess amplitude defined in Eq.\
 \req{eq:sub-amp}.\\

Next, we are going to discuss the amplitude \req{eq:amplitude-AB} in combination with the subprocess
amplitudes \req{eq:amplitude-qg} and \req{eq:amplitude-qq}. 
With the help of \req{eq:property-qg} one can simplify the quark-gluon contribution to the
amplitude \req{eq:amplitude-AB}:
\ba
\lefteqn{\int_{-1}^1 dx_1\int_0^1dx_2 H_A^a(x_1,\xi_1,t_1)\, H_B^g(x_2,\xi_2,t_2)\,{\cal H}_{0a}^{qg}(x_1,\xi_1,x_2,\xi_2)
                                             \=} \hspace*{0.25\tw}   \nn\\
 && \int_0^1dx_1\int_0^1dx_2\Big[H_A^a(x_1,\xi_1,t_1) + H_A^a(-x_1,\xi_1,t_1)\Big]\,   \nn\\[0.01\tw]
 &&       \hspace*{0.15\tw}        \times    H_B^g(x_2,\xi_2,t_2)\,{\cal H}_{0a}^{qg}(x_1,\xi_1,x_2,\xi_2)
\ea
and analogously for the gluon-quark contribution. For the quark GPDs the combinations ($j=A,B$)
\be
H_j^{a(\pm)}(x,\xi,t)\= H_j^a(x,\xi,t) \mp H_j^a(-x,\xi,t)
\label{eq:C-combination}
\ee
are even and odd under the replacement of $x$ by $-x$:
\be
H_j^{a(\pm)}(-x,\xi,t) \=  \mp H_j^{a(\pm)}(x,\xi,t)\,.
\label{eq:symmetry}
\ee
Because of
\be
H_j^{\bar{a}}(x,\xi,t)\=-H_j^a(-x,\xi,t)
\label{eq:antiquark}
\ee
it is obvious that the plus combination in \req{eq:C-combination} which corresponds to the exchange
of a charge conjugation even object in the $t$-channel, refers to a sea contribution
whereas the minus one, corresponding to charge conjugation odd, is a valence-quark contribution \ci{diehl03}.

For the quark-quark subprocess we have to take care of charge-conjugation invariance. Since the photon
in the final state has $C=-1$, we need at one of the hadron vertices in Fig.\ \ref{fig:double-handbag}
$C=-1$ and at the other one $C=+1$. In other words we have to consider the GPD products
$H_A^{a(\pm)}(x_1,\xi_1,t_1)H_B^{a(\mp)}(x_2,\xi_2,t_2)$. The corresponding integral reads
(dropping for a moment the arguments except of $x_i$, for convenience)
\ba
I&=&\int_{-1}^1dx_1\int_{-1}^1dx_2\Big[H_A^{a(-)}(x_1)\,H_B^{a(+)}(x_2)    \nn\\
            && \hspace*{0.2\tw} + H_A^{a(+)}(x_1)\,H_B^{a(-)}(x_2)\Big]\,{\cal H}^{qq}_{0a}(x_1,x_2)\,. 
\ea
The symmetry relation \req{eq:symmetry} allows to write this integral as
\ba
I&=& \int_0^1dx_1\int_0^1dx_2 \left\{\Big[H_A^{a(-)}(x_1)\,H_B^{a(+)}(x_2) + H_A^{a(+)}(x_1)\,H_B^{a(-)}(x_2)\Big]
                                                         \right.\nn\\
    && \left. \hspace*{0.1\tw} \times\, \Big[{\cal H}^{qq}_{0}(x_1,x_2) - {\cal H}^{qq}_{0}(-x_1,-x_2)\Big] \right.\nn\\
    && \left. \hspace*{0.1\tw}  + \;  \Big[H_A^{a(-)}(x_1)\,H_B^{a(+)}(x_2)
                                                    - H_A^{a(+)}(x_1)\,H_B^{a(-)}(x_2)\Big]\right.\nn\\
     &&\left. \hspace*{0.1\tw} \times\, \Big[{\cal H}^{qq}_0(-x_1,x_2) - {\cal H}^{qq}_0(x_1,-x_2)\Big] \right\}\,.
\ea
Using \req{eq:antiquark}, one can further show that
\ba
\lefteqn{H_A^{a(-)}(x_1)H_B^{a(+)}(x_2) + H_A^{a(+)}(x_1)H_B^{a(-)}(x_2) \=} \hspace*{0.25\tw} \nn\\
                &&     2 H_A^a(x_1) H_B^a(x_2) - 2 H_A^{\bar{a}}(x_1) H_B^{\bar{a}}(x_2)\,, \nn\\
\lefteqn{H_A^{a(-)}(x_1)H_B^{a(+)}(x_2) - H_A^{a(+)}(x_1)H_B^{a(-)}(x_2) \=}  \hspace*{0.25\tw} \nn\\
                &&     2 H_A^a(x_1) H_B^{\bar{a}}(x_2) - 2 H_A^{\bar{a}}(x_1) H_B^{a}(x_2)\,.
\ea
We see that the first combination of the GPDs refers to quark-quark and antiquark-antiquark
scattering, i.e.\ it corresponds to Feynman graphs of the type shown on the left-hand side of Fig.
\ref{fig:graphs-qq} whereas the second combination represents quark-antiquark and antiquark-quark scattering
(corresponding to graphs of the type shown on the right-hand side of Fig.\ \ref{fig:graphs-qq}. Thus,
only the type of graphs shown on the left-hand side of Fig.\ \ref{fig:graphs-qq} is to be taken into account.
Quark-antiquark graphs, shown on the right-hand side of this figure, are already included because
of the integrations from -1 to 1.

Putting all together what we have just discussed the amplitude \req{eq:amplitude-AB}
can be cast into the form
\ba
   {\cal M}^{AB}_{\mu_a\mu_b0,\mu_a\mu_b} &=& e_0 \sqrt{1-\xi_1^2}\sqrt{1-\xi_2^2}\,
                         \sum_ae_a \int_0^1dx_1\int_0^1dx_2\nn\\
       &\times& \left\{\Big[ H_A^{a(-)}(x_1,\xi_1,t_1)\,H_B^g(x_2,\xi_2,t_2)\,{\cal H}_{0}^{qg}(x_1,\xi_1,x_2,\xi_2)
                       \right.\nn\\
       &&\left. + H_A^g(x_1,\xi_1,t_1)\, H_B^{a(-)}(x_2,\xi_2,t_2)\, {\cal H}_{0}^{qg}(x_2,\xi_2,x_1,\xi_1) \Big]
                        \right. \nn\\
       &+& \left. \hspace*{-0.02\tw} 2\Big[H_A^a(x_1,\xi_1,t_1)\,H_B^a(x_2,\xi_2,t_2)
                          - H_A^{\bar{a}}(x_1,\xi_1,t_1)\,H_B^{\bar{a}}(x_2,\xi_2,t_2)\Big]\, \right.\nn\\
               &\times&\left. {\cal H}_{0}^{qq}(x_1,\xi_1,x_2,\xi_2) \right. \nn\\ 
         &+& \left.\hspace*{-0.02\tw} 2\Big[H_A^a(x_1,\xi_1,t_1)\,H_B^{\bar{a}}(x_2,\xi_2,t_2)
                          - H_A^{\bar{a}}(x_1,\xi_1,t_1)\,H_B^{a}(x_2,\xi_2,t_2)\Big]\, \right.\nn\\
          &\times& \left. {\cal H}_{0}^{qq}(-x_1,\xi_1,x_2,\xi_2) \right\}\,.
\label{eq:AB-amplitude}
\ea

There are four amplitudes for proton-proton and proton-antiproton collisions for the
helicities $\mu_a=\pm 1/2$ and $\mu_b=\pm 1/2$, for pion-proton collisions only two. Evidently
these amplitudes are the same, there is only one independent amplitude for each process
\be
   {\cal M}^{pp(\bar{p})} \= {\cal M}^{pp(\bar{p})}_{++0,++}\,.
\ee
and analogously for pion-proton scattering.

\section{The electromagnetic lepton-pair production}
\label{sec:elm}
\begin{figure}[ht]
\centering
\includegraphics[width=0.4\tw]{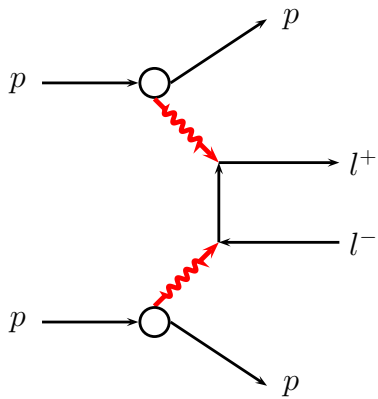}
\caption{\label{fig:elm} Graph for the electromagnetic lepton-pair production in exclusive proton-proton
collisions. To leading-order there is a second graph with the lepton lines crossed. }
\end{figure}

At small $t_i$, the helicity amplitudes for the electromagnetic lepton-pair generation in proton-proton
collisions read (remember the sum of the lepton and antilepton helicity is zero)
\ba
   {\cal M}^{\rm elm}_{\mu_a\mu_b 0,\mu_a\mu_b} &=& \frac{16\pi^2\ale^2}{t_1t_2}\,
                              \frac{G_M(t_1)G_M(t_2)}{\sqrt{(s-s_1)(s-s_2)}}\,
                                \frac{E_q s^{3/2}(Q^2-s_1)(s-s_1)}{s_1^3s_2} \nn\\ 
                       &\times& \Big[ Q^2s_1\sqrt{s} \cos{\theta_q}   \nn\\
                       &+& E_q \big(1-\sin{\theta_q}\cos{\phi_q}\big)
                                  \Big(Q^2 s (1-\cos{\theta_q}) -s_1^2(1+\cos{\theta_q})\Big) \Big]\nn\\
                        &&        + {\cal O}(t_1,t_2)\,.
\label{eq:elm-amp}
\ea                                
The positively charged lepton momentum is defined as
\be
q=E_q(1,\sin{\theta_q} \cos{\phi_q}, \sin{\theta_q} \sin{\phi_q}, \cos{\theta_q})\,.
\ee
Similar contributions exist for proton-antiproton and meson-proton collisions.

The interference between the electromagnetic and the double-handbag contributions is particularly interesting:
It is linear in the GPDs and contains only the real part of the strong amplitude independent on the
regularization scheme exploited. As the electromagnetic and strong amplitudes corresponds to a different
C-parity of the lepton pair, the interference term is antisymmetric with respect to the interchange
of the leptons, in complete analogy to TVCS case \ci{DPB}. As a consequence the interference term
becomes zero if it is integrated over the entire range of dilepton angles.

The interference term may be used as a probe of the handbag  contribution on top of the electromagnetic
one. The theoretically cleanest way would be to consider the differential asymmetry for fixed momenta
of the lepton pair \cite{Pivovarov:2015vya}:
\be
A\=\frac{d\sigma (e^+(p_1) e^-(p_2) )-d\sigma (e^+(p_2) e^-(p_1) )}
                      {d\sigma (e^+(p_1) e^-(p_2) )+d\sigma (e^+(p_2) e^-(p_1) )}\,.
\label{Asym}
\ee
However, this requires  a very high accuracy which can be hardly achieved because of the smallness of
the cross-section. One can also perform integration for polar and/or azimuthal angles, like in TVCS \ci{DPB},
where the spin-dependent and spin-independent terms have different symmetry properties with respect to the
reflection of angles.  This, in turn, would also require a very good acceptance.   
Another method to measure the interference term will be discussed in Sect.\ \ref{sec:pp}. 

\section{Results}
\label{sec:results}
\subsection{Proton-proton collisions}
\label{sec:pp}
For the case of proton-proton collisions we omit the particle labels $A,B$ at the GPDs for
convenience and use the familiar notation $H^a$ for quarks of flavor $a$ and $H^g$ for gluons.
For predictions of the corresponding cross section for lepton-pair production we can make
use of the GPDs extracted from nucleon form factors \ci{DK13} and from electroproduction
of vector mesons \ci{GK3}. In the analysis of the nucleon form factors the GPDs $H$ and $E$
for valence quarks can be extracted for a given parameterization of the zero-skewness GPDs
as a product of the forward limit, the parton densities in the case of $H$, and an exponential
in $t$ with a profile function assumed to be
\be
f_a(x)\= \alpha'_a(1-x)^3\ln(1/x) + B_a(1-x)^3 + A_ax(1-x)^2\,.
\label{eq:profile}
\ee
The parameters $\alpha_a'$, $B_a$ and $A_a$ are fixed from a fit to the nucleon form factor data.
The skewness dependence of the GPDs is generated from the double-distribution ansatz \ci{musatov}
whereby the double distribution is assumed to be a product of the zero-skewness GPD and an
appropriate weight function. The gluon and sea quark GPDs at zero skewness are parameterized
analogously with a small $-t$, small $x$ approximation of the profile function \req{eq:profile}
\be
f_a(x) \simeq \alpha'_a \ln(1/x) +B_a\,.
\ee
For the sea quark GPDs we adopt a result from CTEQ \ci{cteq} it is assumed that
\be
H^{\bar{u}} \= H^{\bar{d}} \= \kappa_s H^s \= \kappa_s H^{\bar{s}}
\label{eq:sea}
\ee
with the flavor-symmetry breaking factor 
\be
\kappa_s\=1 + 0.68/(1+0.52\ln{(Q^2/Q_0^2)}\,.
\ee
With the assumption $H^s=H^{\bar{s}}$ the strange-quark contributions to the amplitude
\req{eq:amplitude-AB} cancel. The initial scale, $Q_0$, for the GPDs is taken as $2\,\gev$.  
The profile functions for the gluon and sea-quark GPDs are obtained from fits to the available
HERA data on $\rho^0$ and $\phi$ electroproduction \ci{h1,zeus}. The parameters of the GPDs can be
found in \ci{GK3}. These GPDs have been used to predict DVCS \ci{KMS} and $\omega$
electroproduction \ci{GK8}; good agreement with experiment is achieved. This strengthens our
confidence in the reliability of the predictions for lepton-pair production. Since in the present
paper we are merely interested in values of $Q^2$ close to the initial scale of $4\,\gev^2$ evolution
of the GPDs plays only a minor role and we simply use their $Q^2$-dependence given in \ci{GK3}. For
$Q^2$ substantially larger than $Q_0^2$ evolution is to be taken into account correctly which, in
principle, can be done with the Vinnikov code \ci{vinnikov}.

\begin{figure}[t]
  \centering
  \includegraphics[height=0.38\tw]{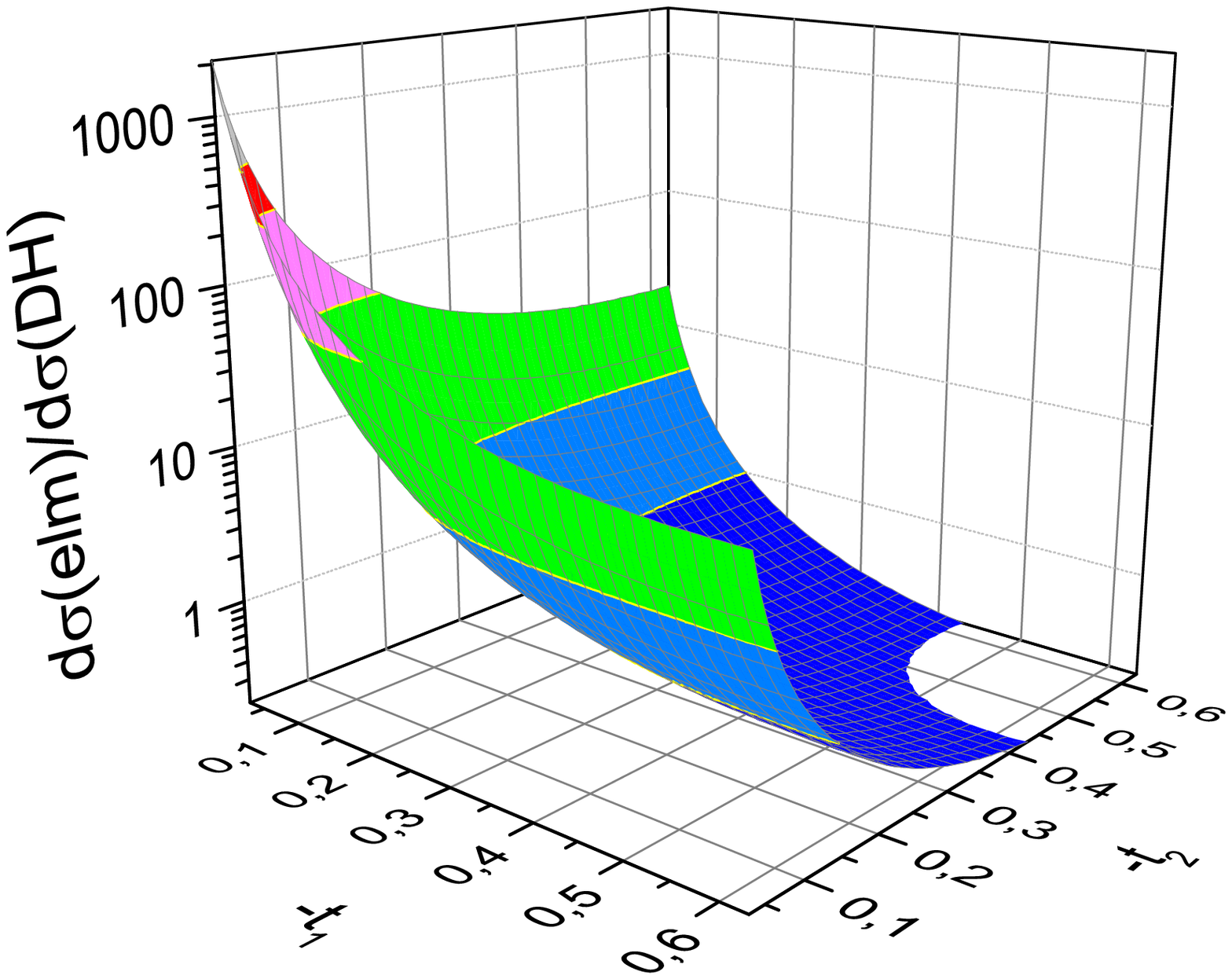}
   \includegraphics[height=0.38\tw]{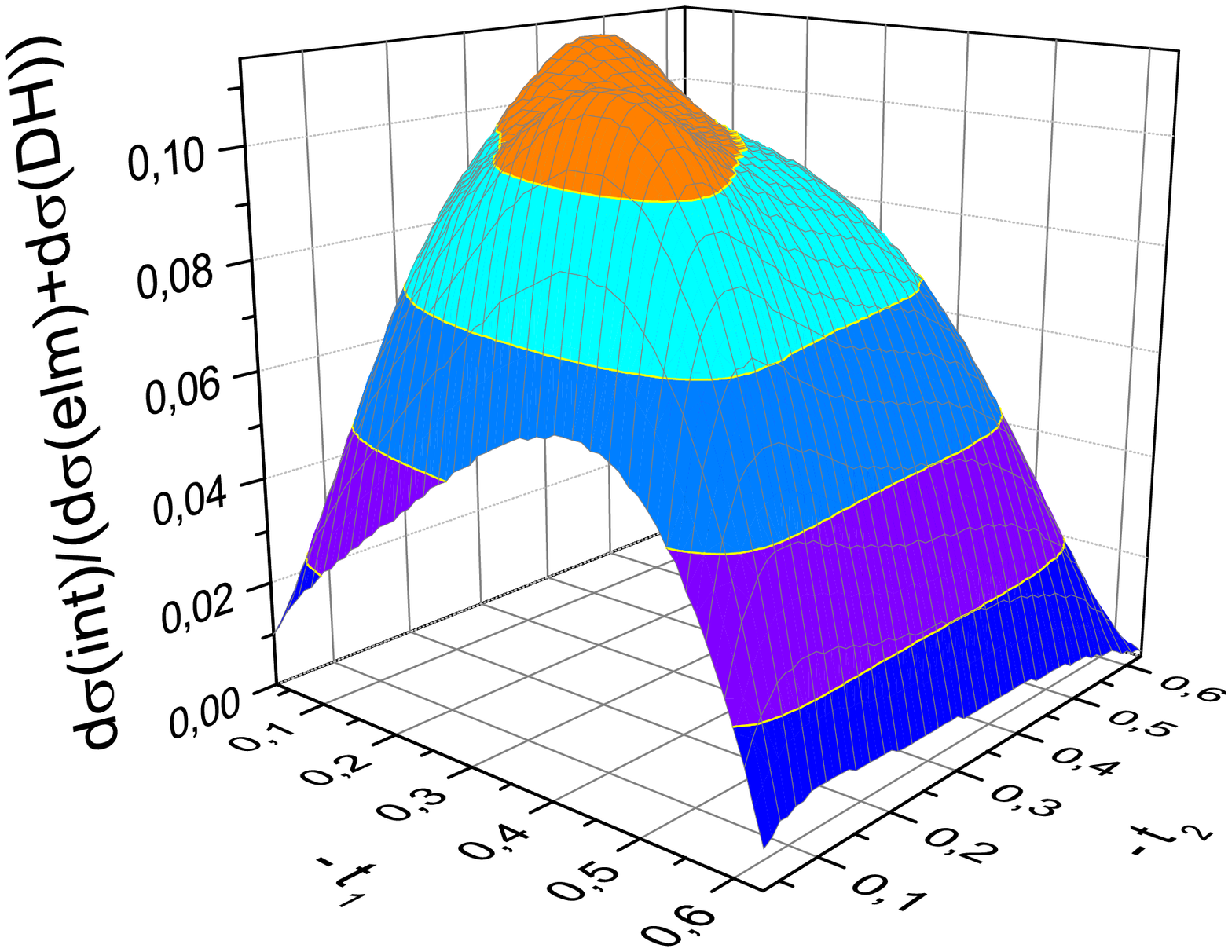}
 \caption{\label{fig:ratio} Left: The ratio of the electromagnetic and DH cross sections for
   $pp\to pp\,l^+l^-$ versus $t_1$ and $t_2$ (in $\gev^2$) at a kinematics typical for NICA
   ($\sqrt{s}=24\,\gev$, $Q^2=3\,\gev^2$).}
 \caption{\label{fig:interf} Right: The ratio of the interference cross section and the sum of the electromagnetic and
   DH contributions at the same kinematics as in Fig.\ \ref{fig:ratio}. The decay angle $\theta^*$ is only
   integrated from 0 to $\pi/2$.}
\end{figure}
The relative strength of the electromagnetic and the DH contributions are displayed in Fig.\ \ref{fig:ratio}
for a typical kinematics accessible at the NICA accelerator. Shown is the ratio of $pp\to pp\,l^+l^-$ cross section
integrated over the full range of dilepton angles (see Eq. \req{eq:cross-section} in App.\ A)
\be
\frac{d\sigma(pp\to pp\,l^+l^-)}{dt_1dt_2dQ^2} \= \frac1{3(4\pi)^5}\,\frac{\ale}{s^2Q^2}\,
\int \frac{ds_1ds_2}{\sqrt{-\Delta_4}}\,|{\cal M}|^2\,.
\label{eq:pp-cross-section}
\ee
where ${\cal M}$ is either the electromagnetic amplitude \req{eq:elm-amp} or the DH one, Eq.\ \req{eq:AB-amplitude}.
Since it is integrated over the dilepton angles there is no interference between the two contributions. Due to the
singular behavior of the electromagnetic amplitude \req{eq:elm-amp} for $t_i\to 0$ it dominates the process at small $t_i$.
Only for $-t_i$ larger than about $0.4\,\gev^2$ the DH contribution takes the lead. The cross sections
\req{eq:pp-cross-section} are symmetric in $t_1$ and $t_2$.\\ 

As mentioned in Sect.\ \ref{sec:elm} the interference between the electromagnetic and the double-handbag contribution is
interesting since it is proportional to the real part of the double-handbag amplitude
\be
d\sigma ^{\rm int} \propto {\cal M}^{\rm elm} {\rm Re} {\cal M}^{pp}\,.
\ee
As above said the double-handbag amplitude is dominantly real and approximately proportional to the product
$H^q H^g$ at $x_i=\xi_i$ even though the product is integrated over a certain range of the $\xi_i$ implied in the
integration over the $s_i$, see Eq.\ \req{eq:skewness}. As one may see from Fig.\ \ref{fig:interf} the ratio
of the interference cross section and the sum of the electromagnetic and the DH contributions is rather large, of the order
of 0.1 at the NICA kinematics, and has two identical maxima at $(t_1,t_2)=(-0.13\,\gev^2,-0.37\,\gev^2)$ and
$(-0.37\,\gev^2,-0.13\,\gev^2)$. In order to obtain a non-zero result for the interference term the photon decay
angle $\theta^*$ (see App.\ A) is only integrated from 0 to $\pi/2$.

\begin{figure}[t]
\centering
\includegraphics[height=0.38\tw]{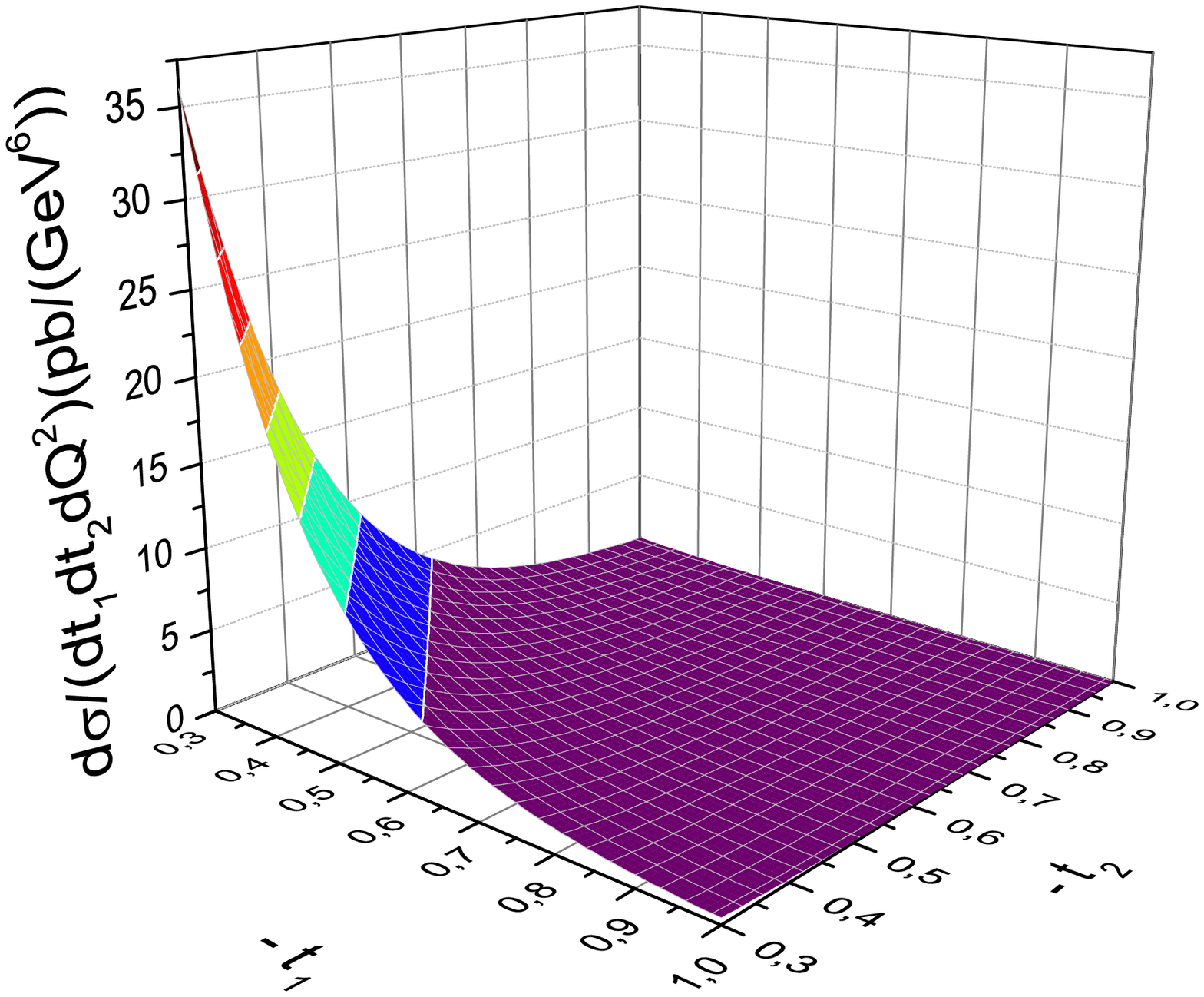}
\includegraphics[height=0.38\tw]{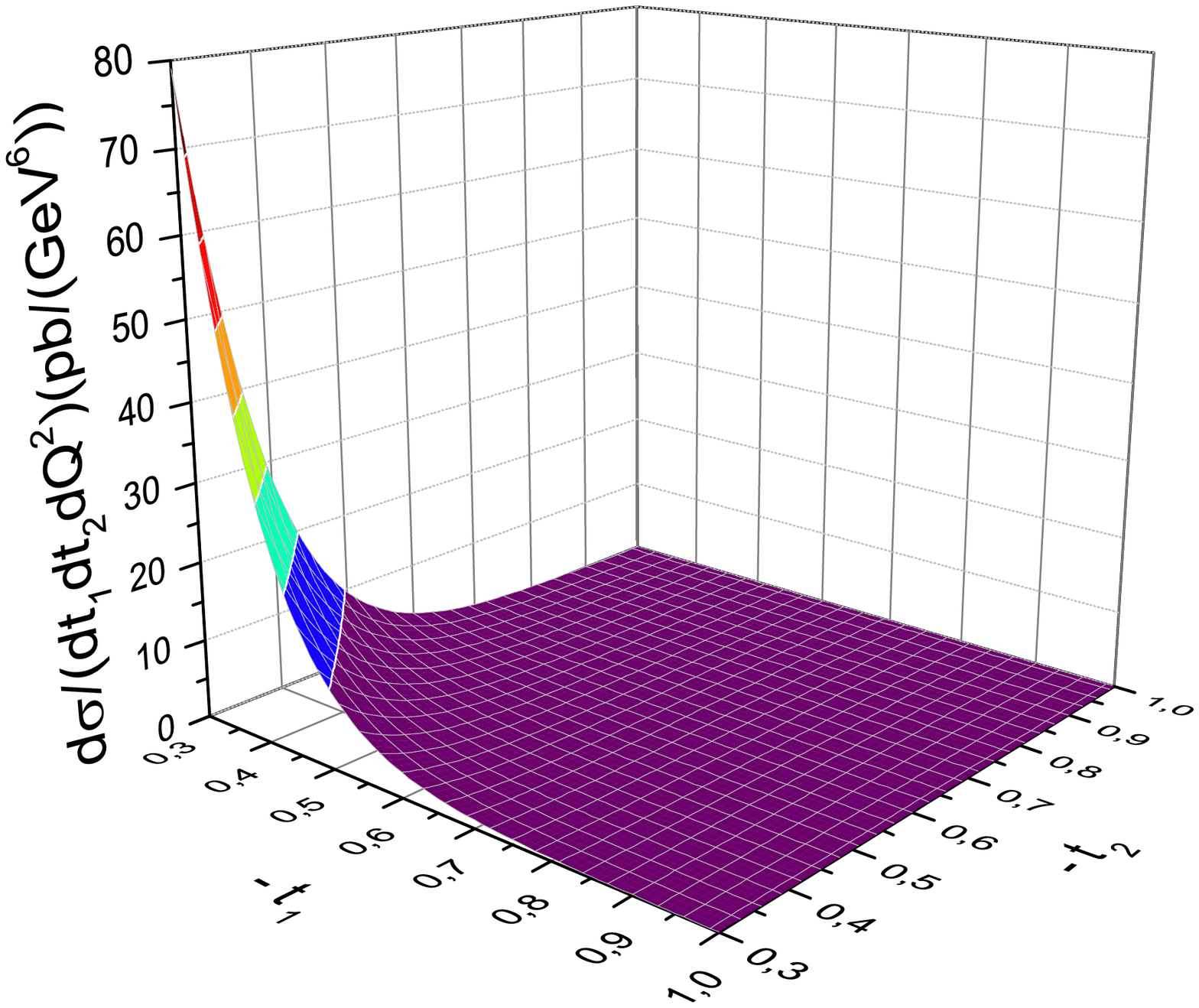}
\caption{\label{fig:nica} Left: The DH contribution to the $pp\to pp\,l^+l^-$ cross section
   (in pb/$\gev^6$) versus $t_1$  and $t_2$ (in $\gev^2$) at $\sqrt{s}=24\,\gev$ and $Q^2=3\,\gev^2$.}
\caption{\label{fig:LHC} Right: The DH contribution to the $pp\to pp\,l^+l^-$ cross section
    in pb/$\gev^6$ versus $t_1$ and $t_2$ (in $\gev^2$) at a typical LHC kinematics:
    $\sqrt{s}=13\,{\rm TeV}$ and $Q^2=5\,\gev^2$.}  
\end{figure} 

In Figs. \ref{fig:nica} and \ref{fig:LHC} the differential cross section \req{eq:pp-cross-section} for $pp\to pp\,l^+l^-$
are shown at a typical kinematics accessible at NICA and at the LHC, respectively. The cross section is only shown for $-t_i$
larger than $0.3\,\gev^2$. Only the DH contribution is taken into account in this region; the electromagnetic contribution
is here neglected. The DH contribution is strongly forward peaked but, for $-t_i\leq 0.3\,\gev^2$, it is overwhelmed by the
electromagnetic lepton-pair generation, see Fig.\ \ref{fig:ratio}. We stress that our numerical studies reveal the dominance
of the quark-gluon subprocess, the quark-quark contribution is almost negligible. In fact,
$|{\cal M}^{qq}|/|{\cal M}^{qg}|\leq 0.1$ for the entire kinematical region explored by us. It is also important to realize
that the main contribution of the quark-gluon subprocess is generated from the imaginary parts
of the two vertex functions. Thus, the subprocess amplitude is dominantly real and approximately proportional to the product
$H^q H^g$ at $x_i=\xi_i$. Since $H^g(\xi_i,\xi_i,t_i)$ strongly increases with decreasing skewness the cross section is rising
with $s$  at fixed $Q^2$ (see \req{eq:skewness}).

\subsection{Proton-antiproton collisions}
\begin{figure}[t]
\centering
 \includegraphics[height=0.43\tw]{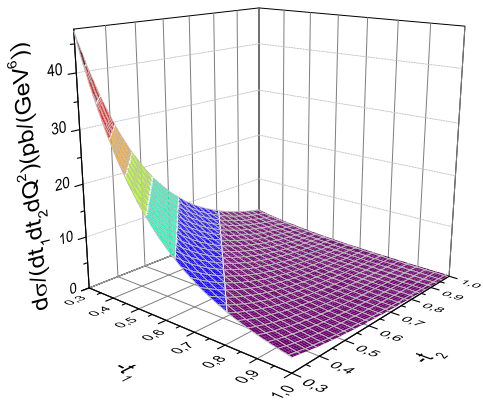}
 \caption{\label{fig:PANDA} The $p\bar{p}\to p\bar{p}\,l^+l^-$ cross section in $pb/\gev^6$
   versus $t_1$  and $t_2$ (in $\gev^2$) at a typical FAIR kinematics: $s=30\,\gev^2$, $Q^2=3\,\gev^2$.}  
\end{figure}

Let us now turn to dilepton production in proton-antiproton collisions which can be measured at the
future PANDA experiment at the FAIR facility. The antiproton GPDs from the lower vertex in Eq.\ \req{eq:AB-amplitude}
are related to the proton ones by
\be
H_{\bar{p}}^{\bar{a}}(x_2,\xi_2,t_2)\= H^a(x_2,\xi_2,t_2)\,, \qquad
                                        H_{\bar{p}}^g(x_2,\xi_2,t_2)\=H^g(x_2,\xi_2,t_2)\,.
\label{eq:p-to-pbar}
\ee
From these relations it is evident that the contributions from the quark-gluon subprocess is the same
for proton-proton and proton-antiproton collisions while the role of the last two terms in \req{eq:AB-amplitude}
are interchanged: The second last term now refers to quark-antiquark scattering
(see the right-hand side of Fig.\ \ref{fig:graphs-qq}) whereas the last one represents quark-quark and
antiquark-antiquark scattering (see left-hand side of Fig. \ref{fig:graphs-qq}. These considerations make
it clear that the cross section for proton-proton and proton-antiproton collisions are identical for the same
kinematics especially since the quark-gluon contribution dominates over the quark-quark one.
In Fig.\ \ref{fig:PANDA} the cross section of the process of interest is displayed for a typical
FAIR kinematics. 

\subsection{Pion-proton collisions}
The last process we want to discuss briefly is $\pi p\to \pi p\,x l^+l^-$. Since the quark-quark contribution
to this process has already been discussed in \ci{pivovarov} we focus our interest to the 
quark-gluon contribution. Applying 
charge conjugation symmetry to the pion GPDs one finds \ci{moutard14}
\be
H_{\pi^+}^a(x_1,\xi_1,t_1)\=-H_{\pi^-}^a(-x_1,\xi_1,t_1)\,.
\ee
In combination with isospin symmetry this leads to (for convenience the variables $\xi_1$ and $t_1$
are dropped for a moment)
\ba
H_{\pi^+}^u(x_1)&=&-H_{\pi^+}^d(-x_1) \= H^{\bar{d}}_{\pi^+}(x_1)\,, \nn\\
H_{\pi^-}^d(x_1)&=&-H_{\pi^-}^u(-x_1) \= H^{\bar{u}}_{\pi^-}(x_1)\,.
\ea
Hence,
\be
H^u_{\pi^\pm}(x_1) \= H^d_{\pi^\mp}(x_1)\,.
\ee
Inspection of the amplitude \req{eq:AB-amplitude} reveals that for the quark-gluon contribution
the sums of the pion's and proton's quark GPDs occur. Each GPD is to be multiplied by the corresponding
quark charge. Because of
\ba
H_{\pi^-}^{u(-)}(x_1) &=& H^u_{\pi^-}(x_1)+ H^u_{\pi^-}(-x_1) \=  H^u_{\pi^-}(x_1)- H^{\bar{u}}_{\pi^-}(x_1) \nn\\
                   &=& H^u_{\pi^-}(x_1)- H^d_{\pi^-}(x_1)
\ea               
and analogously for $H_{\pi^-}^{d(-)}$, the sum of the pion's quark GPD 
simplifies to
\be  
\sum_a e_a H_{\pi^-}^{a(-)}(x_1) \= H^u_{\pi^-}(x_1)-H^d_{\pi^-}(x_2)
\ee
where we made the plausible assumption $H^s_{\pi^-}(x_1)=H^{\bar{s}}_{\pi^-}(x_1)$.
The quark-gluon contribution to the $\pi^-p$ amplitude then reads
\ba
   {\cal M}^{\pi^-p} &=& e_0 \sqrt{1-\xi_2^2} \int_0^1dx_1 \int_0^1 dx_2  \nn\\
   &\times& \left\{\Big[H^u_{\pi^-}(x_1,\xi_1,t_1) - H^d_{\pi^-}(x_1,\xi_1,t_1)\Big] H^g(x_2,\xi_2,t_2)
                     {\cal H}^{qg}_0(x_1,\xi_1,x_2,\xi_2) \right.   \nn\\
  && \left. + H_{\pi^-}^g(x_1,\xi_1,t_1)\sum_a e_a\,H^{a_v}(x_2,\xi_2,t_2)\,
                       {\cal H}^{qg}_0(x_2,\xi_2,x_1,\xi_1)   \right\}\,.
\ea
Thus, only the valence-quark proton GPDs
\be
H^{a_v}(x_2,\xi_2,t_2)\= H^a(x_2,\xi_2,t_2) - H^{\bar{a}}(x_2,\xi_2,t_2)
\ee
contribute.                                  
As in the other cases we investigated, the quark-gluon contribution is much larger than
the quark-quark ones. An analogous result is found for the case of a $\pi^+$ beam.
The generalization of this amplitude to the case of a Kaon beam is straightforward.
Our process $\pi p\to \pi p\, l^+l^-$ as well as $K p\to K p\, l^+l^-$ can be measured at the
future J-PARC accelerator. The measurement of these cross section give in principle access
to the pion and Kaon GPDs. In so far the dynamics can be explored in greater detail than with the
pion (or Kaon) induced exclusive Drell-Yan process \ci{GK9,kumano}.\\
Because of the very limited knowledge of the pion GPDs we refrain from giving numerical estimates
of the $\pi p\to \pi p\,l^+l^-$ cross section.

\section{Summary}
We have investigated lepton-pair production in exclusive hadronic collisions within the handbag
approach. It is assumed that the process amplitude factorizes in a hard partonic subprocess,
$q q (\bar{q})\to qq(\bar{q})\,l^+l^-$ and $qg\to qg\,l^+l^-$, and soft hadronic matrix
elements, $A\to A$ and $B\to B$, which are parameterized as GPDs. We have derived the amplitudes for
this DH mechanism and discussed their properties. The dominant contribution comes from the GPD $H$
in combination with the $qg \to qg\gamma^*$ subprocess. The $qq(\bar{q})\to qq(\bar{q})\gamma^*$
subprocess is also considered but its contribution is much smaller than that from the quark-gluon
one. We have made predictions for the lepton-pair production in exclusive proton-proton and
proton-antiproton collisions for kinematics accessible at NICA, LHC and FAIR.

The DH contribution competes with the purely electromagnetic lepton-pair production. The latter one
is singular for $t_i\to 0$ and, hence, dominates for $-t_i\lsim 0.4\,\gev^2$. The interference between
the two contributions is interesting because it is proportional to the real part of the DH amplitude
which itself is approximately given by the product $H^q H^g$ at $x_i=\xi_i$. However, the interference
term is zero if it is integrated over the entire range of dilepton angles.

We have also briefly examined lepton-pair production in exclusive pion-proton collisions. Measurements
of this cross section which is in principle possible at the future J-PARC accelerator, would
give access to the pion GPDs. The generalization to the corresponding process with a Kaon beam
is straightforward.


{\it Acknowledgements} We are grateful to Helmut Koch for information about the
measurability of $p\bar{p}\to p\bar{p}\,l^+l^-$ with the PANDA experiment.
\begin{appendix}
\section{Phase space and the decay of the virtual photon}
Denoting the lepton momenta by $q$ and $q'$ and inserting the relation
\be
1\= \delta^{(4)}(p_a+p_b-q_1-q_2-q_3)\,d^3q_3dE_3\,,
\ee
one can write the four-particle phase space as
\be
dLips_4(p_ap_b\to q_1q_2qq')\=dLips_3(p_ap_b\to q_1q_2q_3)\,\frac{2E_3dE_3}{2\pi}\,dLips_2(q_3\to qq')\,.
\label{eq:lips4}
\ee
In terms of the invariants the three-particle phase space reads
\be
dLips_3(p_ap_b\to q_1q_2q_3)\=\frac1{(2\pi)^5}\,\frac{\pi}{16 s}\,\frac{dt_1dt_2s_1s_2}{\sqrt{-\Delta_4}}
\ee
where in the massless case \ci{kajantie}
\ba
\Delta_4&=&\frac1{16}\left\{ sQ^2\Big[s(Q^2-2t_1-2t_2)-2(s_1s_2+2t_1t_2-t_1s_1-t_2s_2)\Big]\right. \nn\\
           &&\left. + t_1^2(s-s_1)^2 + t_2^2(s-s_2)^2 + 2s_1s_2t_1(s-s_1) \right.    \nn\\
           &&\left. + 2s_1s_2t_2(s-s_2)- 2t_1t_2s(s-s_1-s_2) + s_1s_2(s_1s_2+2t_1t_2) \right\}.
\label{eq:delta4}
\ea

The decay of the virtual photon is considered in its rest frame, $\Sigma^*$. Then
\be
dQ^2 \=2 E_3 dE_3
\ee
and
\be
dLips_2(q_3^*\to q^*q'{}^*) \= \frac1{32\pi^2}\,d\cos{\theta^*} d\phi^*\,.
\ee
The angles $\theta^*$ and $\phi^*$  are the decay angles in the frame $\Sigma^*$.
The amplitude for the process $AB\to AB ll$ is given by
\be
T^{AB}_{\mu_a\mu_b\lambda-\lambda,\mu_a\mu_b}\=\frac{e_0}{Q} \bar{u}(q,\lambda)\eps(\nu)\cdot \gamma v(q',-\lambda)\,
  {\cal M}^{AB}_{\mu_a\mu_b 0,\mu_a\mu_b} \,.
\ee
Summing the square of the amplitude $T$ over the final state helicities and averaging those in the initial
state we arrive at
\be
|T_L^{AB}|^2\= 2\sin{\theta^*} \frac{e_0^2}{Q^2} |{\cal M}^{AB}_{+ + 0, + +}|^2\,.
\ee
The spin averaged cross section reads
\be
d\sigma(AB\to ABll) \= 4\pi \frac{\ale}{sQ^2} dLips_4 \sin^2{\theta^*}\,|{\cal M}^{AB}_{+ + 0,+ +}|^2\,.
\ee
Using \req{eq:lips4} and integrating over the decay angles of the virtual photon we arrive at
the differential cross section
\be
\frac{d\sigma(AB\to ABll)}{dt_1dt_2dQ^2} \= \frac1{3(4\pi)^5}\,\frac{\ale}{s^2Q^2}\,
\int \frac{ds_1ds_2}{\sqrt{-\Delta_4}}\,|{\cal M}^{AB}_{++0,++}|^2\,.
\label{eq:cross-section}
\ee
 
\end{appendix}  
 
\end{document}